\documentclass[prl,twocolumn,amsmath,amssymb]{revtex4}

\usepackage{graphicx}
\usepackage{bm}
\usepackage{hyperref}

\begin{document}


\title{Bond and N\'{e}el order and fractionalization in  ground states of easy-plane
antiferromagnets in two dimensions}

\author{Kwon Park}
\author{Subir Sachdev}%
\affiliation{Department of Physics, Yale University, P.O. Box
208120, New Haven CT 06520-8120}

\date{\today}

\begin{abstract}
We describe the zero temperature phases and phase boundaries of an
effective lattice model for frustrated, easy-plane, spin $S=1/2$
quantum antiferromagnets. The model contains states with N\'{e}el
and bond-centered charge order and fractionalization in different
limiting regimes, and we describe how these orders compete at
intermediate coupling. The complex Berry phase terms render this
model unsuitable for direct simulation; however, a duality mapping
leads to a model of interacting current loops with positive
weights, which is then studied by Monte Carlo simulations.
\end{abstract}

\pacs{Valid PACS appear here}
\maketitle

Understanding the zero temperature phase diagram of frustrated,
spin $S=1/2$, quantum antiferromagnets in two dimensions
\cite{pwa} is one of the central problems in the theory of
correlated electron systems. Such an understanding is expected to
prove valuable in a description of the cuprate superconductors.
Moreover, a number of frustrated two-dimensional antiferromagnets
with a non-magnetic ground state have been
discovered\cite{Kageyama,coldea}: these could possibly be doped
with charge carriers in the future, leading to exciting
possibilities for new physics.

Some understanding of the different classes of possible ground
states of insulating antiferromagnets has emerged in recent years.
States can be classified by their possession of one or more
distinct types of `order'\cite{senthil}: magnetic order involving
breaking of spin-rotation invariance (as in a N\'{e}el state),
bond order arising from a spontaneous modulation in the magnitude
of the exchange energy which breaks lattice symmetries (as in
spin-Peierls or `plaquette' states), or `fractionalization'
leading to excitations characteristic of the deconfined phase of a
$Z_2$ gauge theory
\cite{senthil,rsprl2,rodolfo2,wenold,mudry,sondhi}. However,
general principles which place constraints on the possible
co-existence of such orders, on phases which can be separated by a
continuous quantum phase transition, and on the topology of the
phase diagram have so far not been delineated. Some issues along
these lines have been addressed in recent
studies\cite{sp,lannert}, where the problems have typically been
related to strongly-coupled field theories whose properties are
not reliably understood.

It is clear that large scale numerical studies will be required to
address these subtle questions, but these have not been possible
so far, because Monte Carlo simulations are impeded by the complex
weights of the quantum spins. Exact diagonalization and series
expansion studies yield only limited information which cannot
easily describe the vicinity of phase boundaries. One strategy to
overcome these obstacles is to abandon study of the quantum spin
model itself, and to focus instead on effective lattice models
which can more easily explore the different limiting regimes of
the phase diagram\cite{rodolfo}. We will follow such an approach
here. We will consider only quantum antiferromagnets with an
easy-plane plane U(1) symmetry, as that makes them amenable to
duality mappings: the dual representation of our lattice model has
only positive weights, allowing Monte Carlo exploration of the
phase diagram on reasonably large lattices. This will allow us to
present the first results on the phase diagram of a model
containing states whose orders extend over all those noted above.

We will study an effective lattice model defined on a
3-dimensional cubic lattice of sites, $j$, representing a
discretized spacetime with the three directions $x,y,\tau$, the
last being imaginary time. Each site has an angular degree of
freedom, $\theta_j$, with $(-1)^{j_x+j_y} e^{i \theta_j} \sim
\hat{S}_{+j}$, the $S=1/2$ spin raising operator on site $j$; the
prefactor of $e^{i \theta_j}$ is the sublattice staggering
associated with the easy-plane N\'{e}el ordering. In addition,
there is a $Z_2$ gauge field, $s_{j,j+\hat{\mu}}=\pm 1$ residing
on the links of the cubic lattice, where $\mu=x,y,\tau$; this is
an auxilliary field which arises in a number of distinct
derivations of the effective model from the underlying quantum
spin model\cite{senthil,rodolfo2,five,sp}, and including it allows
for an especially simply description of phases with
fractionalization. The partition function controlling these
degrees of freedom is
\begin{eqnarray}
&& {\cal Z} = \sum_{\{ s_{j,j+\hat{\mu}} = \pm 1\}} \int \prod_j d
\theta_j \exp \Biggl( K \sum_{\square} \prod_{\square}
s_{j,j+\hat{\mu}} \nonumber
\\ &&\!\!\!\!\!\!\!\!\!
+ \frac{4}{g} \sum_{j, \hat{\mu}} s_{j,j+\hat{\mu}} \cos \left(
\frac{\Delta_{\mu} \theta_j}{2} \right) - i \frac{\pi}{2} \sum_j (
1 - s_{j,j+\hat{\tau}}) \Biggr), \label{u1}
\end{eqnarray}
where $\Delta_{\mu}$ is a discrete lattice derivative
($\Delta_{\mu} f_j \equiv f_{j + \hat{\mu}} - f_j$ for any $f_j$),
and $\square$ indicates all the elementary square plaquettes of
the cubic lattice. A fundamental property of ${\cal Z}$ is its
invariance under the $Z_2$ gauge transformation $s_{j,j+\hat{\mu}}
\rightarrow \eta_j s_{j,j+\hat{\mu}} \eta_{j+\hat{\mu}}$,
$\theta_j \rightarrow \theta_j + \pi(1-\eta_j)$, where $\eta_j =
\pm 1$ is an arbitrary field (we assume periodic boundary
conditions along the $\tau$ direction). The model is characterized
by two dimensionless coupling constants, $g$, and $K$, and we will
explore its phase diagram in the $g$, $K$ plane. The coupling $g$
controls the propensity to N\'{e}el order in the easy-plane, with
the small $g$ region having magnetic order with $\langle e^{i
\theta_j } \rangle \neq 0$. The value of $K$ is controlled most
easily by ring-exchange terms in the underlying quantum spin
system\cite{bfg,five,vortex}. At $K=0$, we can independently sum
over $s_{j,j+\hat{\mu}}$ on each link, leading to weights
dependent only on $\cos(\Delta_{\mu} \theta_j)$, which describes
bosonic, $S_z = 1$ quanta of $e^{i \theta_j}$ hopping in
spacetime; conversely at large $K$ (strong ring exchange), the
$Z_2$ flux of the gauge field $s_{j,j+\hat{\mu}}$ is suppressed,
and the $1/g$ term in (\ref{u1}) describes the propagation of $S_z
= 1/2$ quanta, $e^{i \theta_j /2}$ (`spinons'), allowing the
possibility of fractionalized phases. The last term in (\ref{u1})
is the crucial remnant of the Berry phases of the $S=1/2$ spins of
the quantum antiferromagnet: it imposes quantization of
half-integral spin on each lattice site. Notice that this term can
make the weights in (\ref{u1}) negative: this is the central
reason for the novelty and difficulty of the theory of
two-dimensional quantum antiferromagnets. The theory ${\cal Z}$
was proposed as an effective model for $S=1/2$ quantum
antiferromagnets in Ref.~\onlinecite{five}; it was also
obtained\cite{sp} by taking the easy-plane limit of the effective
lattice models of quantum antiferromagnets with SU(2) symmetry
proposed some time ago \cite{rodolfo}. Upon identifying the spin
raising operator with a hard-core boson $\hat{S}_{+j} \sim
b^{\dagger}_j$, we can also consider ${\cal Z}$ as a model of a
boson system with off-site interactions and plaquette exchange
terms, at a mean density per site which is
half-integral\cite{troyer}.

Sedgewick {\em et al.\/}\cite{sss} have recently studied the
properties of ${\cal Z}$ but without the Berry phase term: their
physical motivation was slightly different (the $e^{i \theta_j}$
represented Cooper pair quanta in a superconductor), for which
this neglect could be justified under suitable conditions. For our
purposes of using ${\cal Z}$ as a model of quantum
antiferromagnets, it is essential to include the Berry phase, and
a description of its consequences is one of the purposes of this
paper. However, because of the non-positive-definite weights, we
cannot perform a direct Monte Carlo sampling over $\theta_j$ and
$s_{j,j+\hat{\mu}}$ configurations, which was the method used by
Sedgewick {\em et al.\/}\cite{sss}. Instead, we will now present a
dual representation of ${\cal Z}$ which does have
positive-definite weights.

The duality mapping\cite{sp} proceeds by first replacing each
cosine term in ${\cal Z}$ by a periodic Gaussian:
\begin{equation}
W(\theta) = {\sum_{\{m_{j \mu}\}}}^{\prime} \exp \left(-
\frac{1}{2g} \sum_{j,\hat{\mu}} \left( \Delta_{\mu} \theta_j - 2
\pi m_{j \mu} \right)^2 \right), \label{villain}
\end{equation}
where the $m_{j \mu}$ are integers on the links of the direct
lattice, and the prime indicates the constraint $m_{j,\mu}$ is
even (odd) for $s_{j,j+\hat{\mu}} = +1 (-1)$. In this form ${\cal
Z}$ can be subjected to a series of mappings which are described
in Ref.~\onlinecite{sp}: these are standard in the theory of
duality of models with $Z_2$ and U(1) symmetry and will not be
reproduced here. The end result is the partition function
\begin{equation}
{\cal Z}_d = \sum_{\{\ell_{\bar{\jmath} \mu}\}} \exp \left( K_d
\sum_{\bar{\jmath}} \varepsilon_{\bar{\jmath}, \bar{\jmath} +
\hat{\mu}} \sigma_{\bar{\jmath}, \bar{\jmath} + \hat{\mu}} -
\frac{g}{8} \sum_{\square} \left( \epsilon_{\mu\nu\lambda}
\Delta_{\nu} \ell_{\bar{\jmath} \lambda} \right)^2 \right)
\label{ud}
\end{equation}
in which $\theta_j$ and $s_{j,j+{\mu}}$ have been integrated out,
and the degrees of freedom are integers $\ell_{\bar{\jmath}\mu}$
on the sites, $\bar{\jmath}$, of the {\em dual} cubic lattice;
$\tanh K_d \equiv e^{-2K}$ is the coupling dual to $K$, the second
sum on $\square$ is over plaquettes on the dual lattice, and
$\epsilon_{\mu\nu\lambda}$ is the antisymmetric tensor. The fields
$\varepsilon_{\bar{\jmath}, \bar{\jmath} + \hat{\mu}}$,
$\sigma_{\bar{\jmath}, \bar{\jmath} + \hat{\mu}}$ are Ising
variables on the links of the dual lattice:
$\varepsilon_{\bar{\jmath}, \bar{\jmath} + \hat{\mu}}$ takes a set
of {\em fixed} values such that $\prod_{\square}
\varepsilon_{\bar{\jmath}, \bar{\jmath} + \hat{\mu}}$ is -1 on
every spatial plaquette and +1 on all other plaquettes (these
values are linked to the Berry phase in (\ref{u1})), while
$\sigma_{\bar{\jmath}, \bar{\jmath} + \hat{\mu}}$ measures the
parity of $\ell_{\bar{\jmath} \mu}$:
\begin{equation}
\sigma_{\bar{\jmath},\bar{\jmath}+\hat{\mu}} \equiv 1 - 2 (
\ell_{\bar{\jmath}\mu} \mbox{ mod 2}). \label{u15}
\end{equation}

The physical meaning of the fields in (\ref{ud}) becomes clear
upon relating them to observable properties of the
antiferromagnet. The quantity $(1/2) \epsilon_{\mu\nu\lambda}
\Delta_{\nu} \ell_{\bar{\jmath} \lambda}$ is associated with links
of the direct lattice, and is the conserved 3-current of the
spin-flip bosons $b^{\dagger}_j$ introduced earlier: notice that
this current can take half-integral values, representing the
transfer of $S_z=1/2$ spinons in a singlet background. Also, the
$\mu=\tau$ component measures the boson density, or
$\hat{S}_{zj}$. The first term in (\ref{ud}) lies on the links of
the dual lattice, each of which can be associated with a plaquette
of the direct lattice; for $\mu=x,y$, each such plaquette is
associated with a single bond of the antiferromagnet. In this
manner we conclude that $\varepsilon_{\bar{\jmath}, \bar{\jmath} +
\hat{\mu}} \sigma_{\bar{\jmath}, \bar{\jmath}+\hat{\mu}}$ for
$\mu=x,y$ is a measure of the exchange energy of the
antiferromagnet on corresponding bonds in the $y,x$ directions.

We begin our study of ${\cal Z}_d$ by considering various limiting
regimes where it reduces to models studied earlier. ({\em i\/})
\underline{$K_d = \infty$ ($K=0$)}: In this case
$\sigma_{\bar{\jmath}, \bar{\jmath} +
\hat{\mu}}=\varepsilon_{\bar{\jmath}, \bar{\jmath} + \hat{\mu}}$
and (\ref{ud}) becomes exactly equivalent to a dual current loop
model of the Hubbard model of the $b^{\dagger}_j$ bosons at
half-integral filling considered by Otterlo {\em et al.}
\cite{otterlo}. This also in keeping with the our earlier
discussion that this limit of (\ref{u1}) describes the facile
motion of $e^{i \theta_j}$ quanta. From the earlier work
\cite{otterlo} we know that the boson ground state is superfluid
({\em i.e.} the spin system has easy-plane N\'{e}el order) for all
$g$ in such a model with only on-site interactions. ({\em ii\/})
\underline{$K_d = 0$ ($K=\infty)$}: The term involving
$\sigma_{\bar{\jmath}, \bar{\jmath} + \hat{\mu}}$ disappears, and
(\ref{ud}) becomes a model of half-boson current loops with
on-site interactions. This is dual \cite{otterlo} to the quantum
XY model of half-angles, $\theta_j/2$, obtained in (\ref{u1}) in
this limit. This model has a superfluid phase at small $g$, and a
fractionalized phase with freely propagating and gapped $S_z =
1/2$ spinons at large $g$. ({\em iii\/}) \underline{$g=0$}: Both
(\ref{u1}) and (\ref{ud}) become trivial, and the ground state has
superfluid (easy-plane N\'{e}el order). ({\em iv\/})
\underline{$g=\infty$}: The action (\ref{u1}) is that of a pure
$Z_2$ gauge theory with a Berry phase term. In (\ref{ud}), this
limit implies that $\sigma_{\bar{\jmath}, \bar{\jmath} +
\hat{\mu}}=\varsigma_{\bar{\jmath}} \varsigma_{\bar{\jmath} +
\hat{\mu}}$ with $\varsigma_j=\pm 1$ an Ising spin on the sites of
the dual lattice; then (\ref{ud}) becomes an 2+1 dimensional Ising
model with every spatial plaquette frustrated, introduced in
Ref.~\onlinecite{rodolfo2} as a model for the non-magnetic states
of frustrated quantum antiferromagnets. The $\varsigma_j$ quanta
are the `visons' of Ref.~\onlinecite{senthil}. This model has a
fractionalized phase at small $K_d$ (visons gapped), and a
confining, bond-ordered phase at large $K_d$ (visons condensed).

\begin{figure}
\centerline{\includegraphics[width=3.2in]{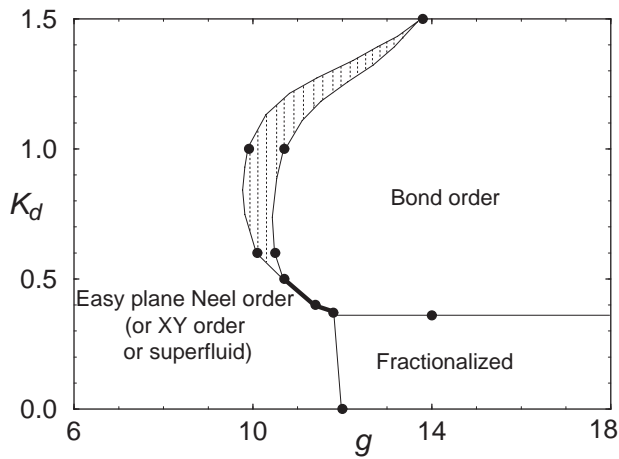}}
\caption{Monte Carlo results for the phase diagram of ${\cal Z}_d$
as function of $g$ and $K_d$. Thin lines represent second order
transitions, while the thick line appears to be first order. We
present evidence below that the hatched region has co-existing
N\'{e}el and bond order. As $K_d \rightarrow \infty$, the region
with superfluid order broadens and eventually extends to all $g$
at $K_d = \infty$. } \label{phase}
\end{figure}

Combining all the limiting cases above, and with information
gained from Monte Carlo simulations on (\ref{ud}) to be described
below, we obtain the phase diagram shown in Fig~\ref{phase}.
The simulations were performed by the Metropolis algorithm on
cubic systems of size $L\times L\times L$, with periodic boundary
conditions.

\begin{figure}
\centerline{\includegraphics[width=2.9in]{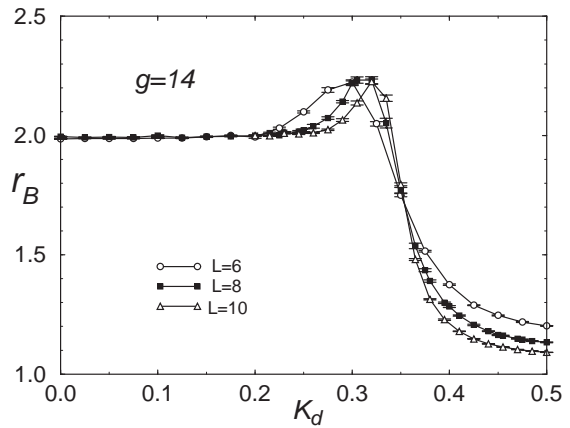}}
\caption{Monte Carlo results for the Binder cumulant, $r_B$, of
the bond order parameter $\Psi_B$ at $g=14$. The results indicate
a second order transition from a phase with $\langle \Psi_B
\rangle = 0$ at small $K_d$ (where the value of $r_B$ is that
associated with a Gaussian random field $\Psi_B$ with zero mean),
to a phase with $ \langle \Psi_B \rangle \neq 0$ at larger $K_d$
(where $r_B=1$, the value expected by $\Psi_B$ condenses). The
crossing points indicate that this transition occurs at $K_d
\approx 0.35$.} \label{dualg14}
\end{figure}

Let us consider first the behavior as a function of $K_d$ for
large $g$. At $g=\infty$, as mentioned above, there is a second
order transition\cite{rodolfo2} from a fractionalized phase at
small $K_d$ to bond-ordered phase at large $K_d$. Our simulations
show that this transition persists into the region with $g$ large
but finite. As before\cite{rodolfo2}, we define a complex
bond-order parameter $\Psi_B = \sum_{\bar{\jmath}}  \left[
(-1)^{\bar{\jmath}_x} \varepsilon_{\bar{\jmath}, \bar{\jmath} +
\hat{x}} \sigma_{\bar{\jmath}, \bar{\jmath}+\hat{x}} + i
(-1)^{\bar{\jmath}_y} \varepsilon_{\bar{\jmath}, \bar{\jmath} +
\hat{y}} \sigma_{\bar{\jmath}, \bar{\jmath}+\hat{y}} \right]$;
this order parameter is non-zero for both columnar and plaquette
orderings of the bonds, which are distinguished\cite{sp} by
distinct values of $\arg( \langle \Psi_B \rangle )$. We merely
tested for non-zero values $\langle \Psi_B \rangle$ and did not
perform the more subtle tests required to determine $\arg( \langle
\Psi_B \rangle )$; this we did by computing the Binder cummulant
$r_B = \langle | \Psi_B |^4 \rangle/\langle |\Psi_B|^2 \rangle^2$.
The results are shown in Fig~\ref{dualg14}, and lead to the
roughly horizontal phase boundary between the bond-ordered and
fractionalized phases in Fig~\ref{phase}.

\begin{figure}
\centerline{\includegraphics[width=3.0in]{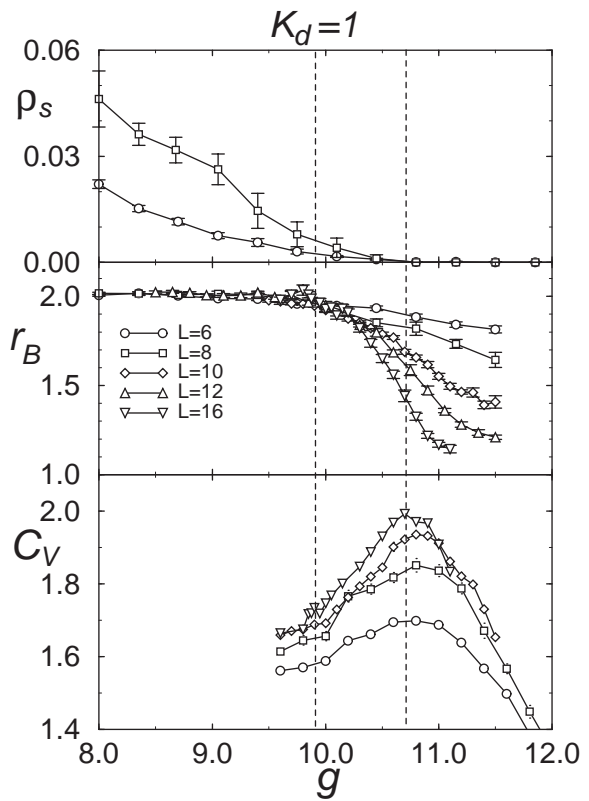}}
\caption{Monte Carlo results for $K_d=1.0$ as a function of $g$.
The dashed lines indicate positions of proposed phase transitions.
Notice that there is a small peak in $C_V$ for the largest system
sizes at the transition at the smaller value of $g$. Data for
$\rho_s$ are not available for the largest system sizes because of
long equilibration times in the winding number sector.}
\label{dualh10cv}
\end{figure}

\begin{figure}
\centerline{\includegraphics[width=2.8in]{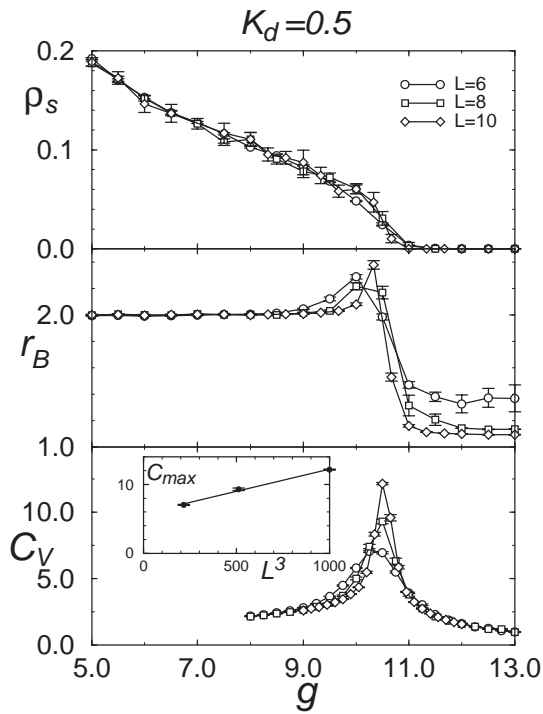}}
\caption{As in Fig~\protect\ref{dualh10cv} but for $K_d = 0.5$.
The inset shows the system volume dependence of maximum value of
$C_V$ at the peak as a function of $g$. The proportionality is
suggestive of a first order transition. Notice that the $L$
dependence of the peak in $C_V$ is much stronger than that in
Fig~\protect\ref{dualh10cv}.} \label{dualh05cv}
\end{figure}

Next we looked at the boundary between the N\'{e}el and
bond-ordered phases as a function of $g$ for different values of
$K_d$. The simplest mean-field theory of ${\cal Z}_d$
predicted\cite{sp} that this should be a direct second-order
transitions. Typical sets of our numerical results are shown in
Figs~\ref{dualh10cv} and~\ref{dualh05cv}.
Apart from $r_B$, we measured two additional physical quantities.
The quantity $C_V$ is the root-mean-square fluctuation in the
action per unit volume: if we interpret ${\cal Z}_d$ in (\ref{ud})
as a classical statistical mechanical model in 3 dimensions, then
$C_V$ would be its specific heat (note, however, that $C_V$ is not
the specific heat of the quantum antiferromagnet). It is an
unbiased measure of the location of phase transitions and of their
order. In addition we measured the stiffness, $\rho_s$ of the
N\'{e}el (or superfluid) order: this was done, as in
Ref.~\onlinecite{otterlo}, by generalizing the current loop model
to its non-zero `winding number' sector, and measuring the
fluctuations in winding number.

The weight of the evidence in Fig~\ref{dualh10cv} for $K_d = 1.0$
favors two second order transitions as a function of $g$, which
are denoted by the dashed lines. At small $g$ we are clearly in
the superfluid phase with a non-zero $\rho_s$, and no bond order.
Conversely at large $g$, bond order is present, but there is no
superfluidity as $\rho_s$ is vanishingly small. However the onset
of superfluid order (as measured by $\rho_s$) clearly occurs at a
point distinct from the onset of bond order (as measured by
$r_B$): there is an intermediate range where superfluid and bond
order are both present. This is also supported by the measurements
of $C_V$, which show peaks at both transitions.

The nature of the data in Fig~\ref{dualh05cv} at $K_d = 0.5$ is
quite distinct. Now the crossovers in $\rho_s$ and $r_B$ happen at
roughly the same value of $g$. Moreover there is only a single
peak in the specific heat, whose maximum value is proportional to
the system volume: this latter observation suggests a first-order
transition.

This paper has described the phase diagram of a 2+1 dimensional XY
model coupled to a $Z_2$ gauge field with a Berry phase term. We
reviewed arguments \cite{senthil,sp,rodolfo,five,vortex} asserting
that this is an effective model for frustrated, easy-plane,
$S=1/2$, two-dimensional quantum antiferromagnets. We believe this
is the first study of quantum phase transitions in the presence of
Berry phase terms (which completely accounts for fluctuations of
gauge fields), and which allows for the various orders that may be
present in the ground state: magnetic, bond, and
fractionalization. We found that fluctuations induced an
intermediate regime with co-existence of magnetic and bond orders
in our U(1) symmetric model; a similar co-existence has also been
argued recently for frustrated $S=1/2$ models with SU(2) symmetry
\cite{sushkov,sp}.

We thank T.~Senthil and M.~P.~A.~Fisher for very useful
discussions. This research was supported by US NSF Grant DMR
0098226.


\vspace{-0.2in}

\begin{thebibliography}{}

\vspace{-0.2in}

\bibitem{pwa} P.~W.~Anderson, Science {\bf 235}, 1196 (1987).

\bibitem{Kageyama} H.~Kageyama {\em et al.}, Phys. Rev. Lett.
{\bf 82}, 3168 (1999).

\bibitem{coldea}  R.~Coldea, D.~A.~Tennant, A.~M.~Tsvelik, and
Z.~Tylczynski, Phys. Rev. Lett. {\bf 86}, 1335 (2001); R.~Coldea
{\em et al.},
\href{http://arxiv.org/abs/cond-mat/0111079}{cond-mat/0111079}.

\bibitem{senthil} T.~Senthil and M.~P.~A.~Fisher, Phys. Rev. B
{\bf 62}, 7850 (2000).

\bibitem{rsprl2} N.~Read and S.~Sachdev, Phys. Rev. Lett. {\bf 66},
1773 (1991).

\bibitem{rodolfo2} R.~Jalabert and S.~Sachdev, Phys. Rev. B {\bf 44}, 686
(1991).

\bibitem{wenold} X.-G.~Wen, Phys. Rev. B {\bf 44}, 2664 (1991).

\bibitem{mudry} C.~Mudry and E.~Fradkin, Phys. Rev. B {\bf 49},
5200 (1994).

\bibitem{sondhi} R.~Moessner, S.~L.~Sondhi, and E.~Fradkin,
\href{http://arxiv.org/abs/cond-mat/0103396}{cond-mat/0103396}.

\bibitem{sp} S.~Sachdev and K.~Park,
\href{http://arxiv.org/abs/cond-mat/0108214}{cond-mat/0108214}.

\bibitem{lannert} C.~Lannert, M.~P.~A.~Fisher, and T.~Senthil,
Phys. Rev. B {\bf 63}, 134510 (2001).

\bibitem{rodolfo} S. Sachdev and R. Jalabert, Mod. Phys. Lett. B {\bf 4}, 1043 (1990).


\bibitem{five} E.~Demler, C.~Nayak, H.-Y.~Kee, Y.~B.~Kim, and
T.~Senthil,
\href{http://arxiv.org/abs/cond-mat/0105446}{cond-mat/0105446}.

\bibitem{bfg} L.~Balents, M.~P.~A.~Fisher, and S.~M.~Girvin,
\href{http://arxiv.org/abs/cond-mat/0110005}{cond-mat/0110005}.

\bibitem{vortex} Using (\ref{villain}), we can also write ${\cal
Z}$ as \protect\cite{sp} ${\cal Z} = \int \prod_j d \theta_j
W(\theta) e^{-S_B - E_v (n_{\bar{\jmath}\mu})}$ where $S_B$
contains the Berry phases of the underlying spins, the sum over
$m_{j\mu}$ is unconstrained (unlike in (\ref{villain})),
$n_{\bar{\jmath}\mu} = \epsilon_{\mu\nu\lambda} \Delta_{\nu}
m_{j\lambda}$ is the current of the vortices in the $b_j$ bosons,
and $E_v$ depends only on $n_v \mbox{(mod 2)}$, with $E_v
(\mbox{even}) - E_v (\mbox{odd}) = -2K$. So $K$ lowers the core
energy of {\em double} vortices, as discussed in
\protect\cite{bfg}.

\bibitem{troyer} F.~Hebert, G.~G.~Batrouni, R.~T.~Scalettar,
G.~Schmid, M.~Troyer, and A.~Dorneich,
\href{http://arxiv.org/abs/cond-mat/0105450}{cond-mat/0105450}.

\bibitem{sss} R.~D.~Sedgewick, D.~J.~Scalapino, and R.~L.~Sugar,
\href{http://arxiv.org/abs/cond-mat/0012028}{cond-mat/0012028}.

\bibitem{otterlo} A. van Otterlo, K.-H.~Wagenblast, R.~Baltin,
C.~Bruder, R.~Fazio, and G.~Sch\"on, Phys. Rev. B {\bf 52}, 16176
(1995).

\bibitem{sushkov}  O.~P.~Sushkov, J.~Oitmaa, and Zheng Weihong,
Phys. Rev. B {\bf 63}, 104420 (2001).

\end{thebibliography}
\end{document}